\documentstyle[12pt,epsf,epsfig]{article}

\def\be{\begin{equation}}
\def\bea{\begin{eqnarray}}
\def\ee{\end{equation}}
\def\eea{\end{eqnarray}}

\begin{document}
 
\title{Exact critical exponent for the shortest-path scaling
function in percolation}
\author{Robert M. Ziff\\
{\sl \small University of Michigan, Ann Arbor MI, USA 48109-2136}}
\maketitle
 
\medskip
\medskip
 
\begin{abstract}
\noindent
It is shown that the critical exponent $g_1$ related to pair-connectiveness
and shortest-path (or chemical distance) scaling, recently studied by
Porto et al., Dokholyan et al., and Grassberger, can be found exactly 
in 2d by using a crossing-probability result of Cardy,
with the outcome $g_1 = 25/24$. This prediction is consistent with existing
simulation results.
[Published as J. Phys. A. 32, L457-459 (1999)]
\end{abstract}


An important quantity describing percolation clusters is
the chemical distance or shortest path \cite{HavlinBenAvraham}.
There has been considerable effort studying its 
scaling properties for distances small compared
with the size of the cluster (i.e., \cite{BarmaRay,Bunde,Roman})
including recent work by
Porto et al.\ \cite{porto}  and Dokholyan et al.\  
\cite{dokholyanJSP,dokholyanPA}.
Very recently, Grassberger \cite{grass99}
has  shown that these scaling properties can 
be analyzed efficiently by studying the growth
of two nearby clusters, a method first suggested
in the work of Dokholyan et al.\ \cite{dokholyanJSP}. In this note I 
show that a scaling relation for the growth of two clusters
\cite{dokholyanJSP} can be combined with  
a previous result of Cardy \cite{cardy98} to find
an exponent of the
shortest-path behavior exactly.

In \cite{grass99}, Grassberger considered the function $N(t)$ (which
I call $N_2(t)$) giving the probability
that two clusters grown from nearby seeds survive at least to time $t$, 
where clusters are grown by a
Leath-type algorithm \cite{leath} and $t$ is the number of generations
or equivalently the chemical distance from the seeds to the growth sites.
(To survive up to that time means that both clusters survive and remain
distinct.) \ 
Another interpretation of $N_2(t)$ is that it gives the probability that
two sites appear to belong to two different infinite clusters,
when the environment is probed up to a chemical distance $t$ from the
two sites.  (As discussed in \cite{grass99}, two points a finite
distance apart in fact belong to two different infinite clusters with
probability zero.)  At the critical point, $N_2(t)$ is presumed to
behave as a power-law
\be 
   N_2(t) \sim t^{-\mu}																			      \label{Nt}
\ee
as $t \to \infty$.
Grassberger also considered the probability $p(t)$ that the two clusters
coalesce {exactly at time $t$; $p(t)$ is proportional
to  $\rho({\bf x-y},2t)$, where ${\bf x}$ and ${\bf y}$ are the locations
of the seed points and
 $\rho({\bf x},t)$ is the pair-connectiveness function, which behaves as 
\cite{grass82,grass92}
\be
   \rho({\bf x},t) \sim {1\over t^{1+2\beta/\nu_t}} \phi(r/t^z) 
\ee
in the scaling limit.
The scaling function $\phi(\zeta)$ is presumed to
behave as 
$\zeta^{g_1}$ for $\zeta \to 0$ \cite{porto}, so
$\rho({\bf x}, t) \sim r^{g_1}$ for constant $t >> r^{1/z}$.
Grassberger showed that these  arguments imply $p(t) \sim
t^{-\lambda}$ with 
\be 
   \lambda = 1+{2\beta\over \nu_t} + z g_1					\label{lambda}
\ee
and furthermore argued that $\mu = \lambda - 1$.  
The relation (\ref{lambda}) was first given 
(in a slightly different
notation) by Dokholyan et al.\
\cite{dokholyanJSP}.

Based upon an analogy to self-avoiding random walks,
Porto et al.\
\cite{porto} conjectured that $g_1$
is related to $d_{\rm min} = 1/z = \nu_t/\nu$ by
\be
   g_1 = d_{\rm min} - \beta/\nu 
\qquad (\hbox{conjecture})           \label{g1conj}
\ee 
which they found to be supported, within the $\approx 5$\% error bars,
by numerical measurements.
This conjecture also appears in \cite{dokholyanJSP,dokholyanPA}.
Inserting eq.\ (\ref{g1conj}) into eq.\ (\ref{lambda}) implies
\be
   \lambda = 2 + {\beta\over \nu_t} 
\qquad (\hbox{conjecture}) \; .        \label{lambdaconj}
\ee
However, from precise simulations of
$p(t)$ and $N_2(t)$,
Grassberger found strong numerical evidence against the above conjecture
(and
provided theoretical arguments against it as well). He found, in 2d,
\be
 \mu = 1.1055(10), \qquad \lambda = 2.1055(10), 
 \qquad g_1 = 1.041(1)                       \label{gvalues} 
\ee
which are numerically inconsistent with the predictions $\mu = 1.09213(5)$
and $g_1 = 1.0264(3)$ that follow from 
eqs.\ (\ref{g1conj}) and (\ref{lambdaconj})
and the known values $\beta = 5/36$, $\nu = 4/3$, and
$d_{\rm min} = 1.1306(3)$ \cite{grass99}, where
numbers in parentheses following
numerical data represent statistical errors in the last digit(s).

Here I show that $g_1$ can be found exactly by
relating $N_2(t)$ to a crossing problem
solved recently by Cardy \cite{cardy98}.  Cardy has shown that
for a rectangular system of dimensions $L_v \times L_h$, 
with periodic boundary conditions in the vertical
direction, the probability of having at least $k$ clusters cross in the
horizontal direction behaves, for large aspect ratio $R = L_h/L_v$, as 
\be
   P_k(R) \sim e^{-a_k R}
 \label{Pk}
\ee
with
$a_1 = 5 \pi / 24$ and $a_k = (2\pi/3)(k^2-1/4)$ for $k > 1$. 
(The formula for $k=1$ is 
different because for one cluster it is not necessary
to also have a crossing cluster on 
the dual lattice, while for $k > 1$ crossing clusters
there must be $k$
crossing dual-lattice clusters.)  The probability
that at least two clusters (or, to the same order,
{\it exactly} two clusters)
cross the rectangle
is given by $P_2 \sim \exp(-5\pi R/2)$.

Crossing problems in critical percolation 
are believed to be conformally invariant, because under a conformal
transformation, in which all elements only expand or contract, 
the crossing
properties of each element should remain unchanged \cite{aizenman,Langlands}.
One can transform the
rectangle into an annulus by putting the four corners of the rectangle at
$z = 0, 2 \pi R, 2 \pi R + 2\pi i$ and $2\pi i$ on the
complex-$z$ plane, and letting $z' = e^z$.
The result on the $z'$-plane is an annulus with an inner radius of
$1$ and an outer radius of $r = e^{2 \pi R}$. \
The top and bottom edges of the rectangle
close together, exactly matching the periodic boundary conditions.
Assuming conformal invariance of the crossing probability,
it follows from (\ref{Pk}) that the probability $p_k$ that at least $k$
clusters cross between the inner and outer boundaries of the 
annulus is given by
\be
 p_k(r) \sim r^{-a_k / (2 \pi)}  
 \label{pk}
\ee
or $p_1(r) \sim r^{-5/48}$, $p_2(r) \sim r^{-5/4}$,
$p_3(r) \sim r^{-35/12}$, etc.  Now, one can associate $p_2(r)$ with
the quantity $N_2(t)$ of eq.\ (\ref{Nt})
by transforming from chemical distance $t$ to the
radial distance $r$ using $r \sim t^z$.   This
yields
 \be
 N_2(t) \sim  t^{-5z/4}         \label{Nt2}
\ee
which implies by (\ref{Nt})
 \be
\mu = {5z\over 4} = {5 \nu_t\over 4 \nu} = 1.1056(3)    \label{muresults}
\ee
and by (\ref{lambda}) gives
\be
\qquad g_1 = {5\over 4} - {2 \beta \over \nu} = {25\over
24} = 1.041666\ldots   \ .                                \label{g1results}
\ee
These predictions are consistent with Grassberger's measurements 
(\ref{gvalues}) as well as Porto et al.'s determination $g_1 = 1.04(5)$. 
However, (\ref{g1results}) is apparently inconsistent with the conjecture 
(\ref{g1conj}}), since it would imply
\be
d_{\rm min} = {5\over 4} - { \beta \over \nu} = {55\over
48} = 1.1458333\ldots \qquad (\hbox{conjecture})     \label{dminresults} 
\ee
which differs from the measured values of
$d_{\rm min}$, 1.1306(3) \cite{grass82}
and 1.130(4) \cite{HerrmannStanley}.

Another way of looking at Grassberger's numerical 
results is that they serve to
confirm the ideas of conformal invariance and Cardy's formula for $k = 2$ 
to high precision.  Note that Cardy's formula for $k=1, 2$ and 3 has also been
verified numerically by Shchur and Kosyakov
\cite{shchur97,shchur98,shchur99}.
Indeed, eq.\ (\ref{Nt2})  can be generalized for the
probability $N_k(t)$ that $k$ clusters remain alive and distinct 
up to time $t$, 
\be
N_k(t) \sim p_k(t^z) \sim t^{-z a_k / (2 \pi)}      \label{Nk} 
\ee
so that $N_3(t) \sim t^{-35 z / 12} \sim t^{-2.580}$ etc.  
Grassberger has also measured this
quantity for $k=3$ and $k=4$, and the behavior he finds is consistent
with the above predictions \cite{grass99pri}.

I note finally that the relation $p_1(r) \sim r^{-5/48}$
following (\ref{pk}) is just the statement 
that the probability a cluster grows to a radius greater
or equal than $r$ is $p_1(r) \sim r^{D-d}$.
The latter formula follows from $P_{\ge s} = \int_s^\infty s n_s \sim s^{2 -
\tau}$ with $s = r^D$ with $D = 91/48$,
and the hyperscaling relation $\tau -
1 = d/D$. Transforming from the annulus to a rectangle
yields Cardy's result
(\ref{Pk}) for $k=1$, 
 $P_1(R) \sim e^{-2 \pi (d - D) R}$ (valid for $d = 2$ only).

In conclusion, I have shown that
the density of growth sites on 2-d percolation clusters behaves
as $r^{25/24}$ for large time and small $r$ . 
 
The author thanks P. Grassberger, J. Cardy and S. Havlin for comments.
This material is based upon work supported by the
U. S. National Science Foundation Grant No.\ DMR-9520700.

\end{document}